**The Creative Process in Musical Composition: An Introspective Account**

Liane Gabora
University of British Columbia

For correspondence:
Dr. Liane Gabora
Department of Psychology
University of British Columbia
Okanagan campus, 3333 University Way
Kelowna BC, V1V 1V7, CANADA
Email: liane.gabora@ubc.ca

I recently reviewed a book that suggested there is a gentle backlash taking place against the shift over the last few decades away from introspective accounts in psychology to experiments in strictly controlled laboratory conditions. It is suggested that a more nuanced, contextual, multifaceted understanding of psychological phenomena might be possible through an amalgamation of experimental and introspective approaches. This gave me a sense of permission to try to put into words the subjective experience of engaging in a particular creative act. The result is something quite different from anything I have previously written for publication on the topic of creativity.

This chapter charts my creative process in the composition of a piece of music titled 'Stream not gone dry' that unfolded, while I was primarily occupied with other matters, over the course of nearly two decades. It avoids any discussion of the technical aspects of musical composition, and it can be read by someone with no formal knowledge of music. The focus here is on what the process of composing this particular piece of music says about how the creative process works. My interpretation of the music-making process may be biased by my academic view of creativity. I believe, however, that the influence works primarily in the other direction; my understanding of how the creative process works is derived from experiences creating. This intuitive understanding is shaped over time by the process of reading scholarly papers on creativity and working them into my own evolving theory of creativity, but the papers that I resonate with and incorporate are those that are in line with my experience. This chapter just makes the influence of personal experience more explicit than in other more scholarly writings on creativity.

I consciously decided to devote my life to creative pursuits and have attempted to do something creative in many different fields, not out of design, but because that's where I was led by the questions I wanted to answer and the feelings I needed to express. Thus I believe I am in a pretty good position to use personal experience as a guide in the development of a theory of creativity. This does not mean that what is written here is an accurate depiction of how the creative process works, or of how the musical composition process works, or even of what went on during the process of writing this particular piece of music, for introspective accounts are notoriously flawed. The introspective account portrayed here is one piece of evidence for a particular way of understanding the creative process amongst a growing body of knowledge that includes studies with human participants and formal mathematical and computational models. One might assume that, for such an introspective account to be of value, it is necessary that the creative outcome be deemed of high quality in the eyes of the world. Few people have heard 'Stream not gone dry' and those who have might say it's good simply because they know me and would not want to hurt my feelings. The very specific comments they give make me think that they genuinely like it, but I have no idea whether the piece would be considered meritorious outside my circle of friends and acquaintances. However, I do not think it is essential that a creative work be deemed by experts to be of high quality to merit a chapter about it; a creative process is of value to the extent that it exerts a transformative effect on the creator (Gabora & Merrifield, 2013; Gabora, O'Connor, & Ranjan, 2012). There is a growing body of evidence that creative practise can indeed be therapeutic (Ball, 2002; Bell & Robbins, 2007; Chambala, 2008; De Petrillo & Winner, 2005; Drake, Coleman & Winner, 2011; Hennessey & Amabile, 2010; Moon, 1999, 2009), including evidence that playing music can be therapeutic (Boothby & Robbins, 2011; Colwell, Davis, & Schroder, 2005; Forinash, 2005; Saarikallio & Erkkila, 2007). From a first-person perspective, the experience of nurturing a raw musical idea into this eight minute long piece of music had a cathartic effect on me. Had I not composed it I would have no real understanding of 'grace' or 'forgiveness' or 'deliverance'; I would be, at least from the inside but I think on the outside too, a very different person.

## History of the Piece of Music

There is a sense in which I believe that the history of any creative work is deeply rooted in the murky past, extending further back in time than the birth of its creator to those who created products and events that influenced the creator, and to those who influenced *these* people, and so forth. This is the case with respect to 'Stream not gone dry'. Not long after I started taking piano lessons as a young child I became aware that whenever I heard a piece of music, part of my brain was figuring out how I would play it on the piano, whether or not there was a piano around. If it wasn't immediately obvious how it would be played, my brain would continue trying to figure it out, for potentially hours, days, or even weeks. This made people think I was nervous because I appeared to be tapping my fingers, and absent minded because I wasn't paying as much attention as I would be otherwise to what was going on around me. The upside, though, is that by the time I actually sat down at a piano, I already had the song largely figured out. When I was young this was extremely arduous; what kept me going was the exhilaration and sense of mastery I felt when I had learned to my satisfaction to play the song by ear. Eventually I started to be annoyed by any music that was so simple that it was instantly obvious how to play it. I became obsessed by music that had tricky things going on, not gratuitous tricky things, but a kind of complexity that launched me into a cathartic or blissful state, such as 'Don't look back' by Boston. I was unable to stop thinking about this piece of music until I had successfully figured out how to play it. I would run back and forth from the dining room where the piano was to the living room playing certain parts of the record (this was a while back) over and over until I had nailed it. When I did nail it, I felt as if the music had revealed its magic to me, and in a sense it had become mine. I think pianists can feel this even more, in a certain sense, than members of the original band or orchestra, because the piano version incorporates multiple parts or instruments at once. The pianist does in parallel with ten fingers what it took several musicians to do, and in this way comes to know the music not just in terms of its individual parts but in terms of how they come together. This without a doubt presents challenges, and perhaps also rewards, that were not faced by the original members of the band or orchestra.

If you constantly have this going on as a background process in your brain all your life, you eventually become fairly decent at it. You go through life constantly making observations to yourself about what kinds of sounds express and evoke particular emotional experiences. The sense of mastery and even feeling of intimacy I developed toward music composed by other people was undoubtedly a necessary prelude to composing my own music.

Despite claiming that the history of 'Stream not gone dry' goes back to my childhood, or even earlier, I would be equally correct to say that it came to me suddenly. I was sitting alone in my bedroom in the early nineties, in the depths of despair. I was dealing with the kind of belittling sexist treatment and abuse of power that many women who enter the sciences have encountered, which left me with a sense of having found myself in the wrong universe. A specific event crystallized these feelings and triggered the following experience. I had a sudden strong sense that a wise female being was hovering just above me, expressing, though not in words, that she cared about me, and knew that I didn't deserve to be treated how I had been treated, and that I deserve to be seen for who I am.

I am not superstitious, and am happy to believe this was all in my head. I was so convinced that I didn't believe in anything like angels that it wasn't until some time later that it even occurred to me to wonder: could *that* be what people mean by an angel? I have never had another such experience, and do not know how to begin to scientifically comprehend or analyse it. So I will not dwell on it further except to say that, although it is cliché to say that creative insight comes at once in a burst of inspiration from a source that is simultaneously outside oneself yet at the core of oneself, that *is* what it felt like for me.

From that moment on I knew the melody of the song, and the bridge, and the words, which I wrote out immediately. However, it was over a year later that I took an electronic music course and finally worked out a piano version that I was reasonably happy with. I knew, however, that it was incomplete, and that I would one day come back to it. It never consciously occurred to me that to fully enter into the spirit of this music again I would have to re-experience events like those that had originally inspired it.

Eventually I left science and found myself doing psychology. One day, many years later, a psychology professor told me that in his previous job he had felt badly that he and several of the male professors he worked with played hockey and then hung out drinking beer several times a week the male graduate students, but the female students were excluded. Then he said that there were other regular exclusively male events that the female students were never invited to because the male professors did not want to risk becoming attracted to them, or risk being accused of pursuing them. The female students knew they were excluded but no one ever explained to they why.

I was so startled to learn that this kind of differential treatment was not limited to the sciences but extended to psychology that I started shaking, and plunged into a state of hopelessness and despair for days, followed by a longer lasting sense of dread and foreboding. This over-reaction made me realize that I had never recovered from the events that had occurred nearly two decades earlier earlier. Phase two of serious work on 'Stream not gone dry' was brought on by events similar to those that had inspired me to begin writing it. Music is highly effective at triggering autobiographical memories (Tillmann, Peretz, & Samson, 2011), and in my experience the reverse is also true: events that push you back into reliving certain experiences can trigger a re-entry into music that was written at that time.

I did not actually *decide* to re-work the music. What initiated it was the following. For at least ten years, whenever I played 'Stream not gone dry', I immediately afterward started playing the beginnings of another composition that was (to me) more complex and tumultuous and intriguing. I had told myself that if I was ever finally satisfied with the first piece I would work on this more complex piece. One day, right as I approached the triumphant climax of 'Stream not gone dry', instead of the triumphant climax, I spontaneously played something that sounded very much like crying, which somehow led directly to playing the more complex piece. The two were connected! I was simultaneously happy to see that they had 'found each other', and sad to realize that I only had one piece after all, not two.

I started to explore this, and before I knew it I was deep into tearing the piece apart so that I could rebuild it. Thus began a little era of my life fraught with stress and excitement, lived in near isolation. I could no longer play the way I had played it for years without it now sounding extremely wrong, yet I could not yet play it properly in its new form, amalgamated to the more complex piece. And there was a sickening sense of things coming together in a way that, while disagreeable in the extreme, made sense to a certain part of my brain. The first time around I hadn't learned certain lessons and healed certain wounds. That, it seemed, was why I had never finished the music, and why the past had come back to haunt me. The part that sounds like crying, the "silence before the storm", is necessary to fully launch the mind into the chaotic frenzy of the complex part, which in turn is needed in order to feel the reassuring calm (the "silence that follows the storm") when the melody quietly returns. In the process of figuring out how best to make the raw emotions that inspired the piece come through in the music, they transformed into something that was (comparatively) objective and technical; the emotions found a form in which they could live in peace with the rest of what exists in the world, both within me and outside of me.

The act of playing the piece itself is a constant reminder of what the piece is about because of how a piano is constructed. There was no point in my ever trying to become a concert pianist, simply because someone long before I was born someone decided that piano keys would be of a width of that is optimal for someone whose hands are much bigger than mine. I can only just reach an octave, and when it is necessary to play several octaves in quick succession as is often the case in this piece, I am much more prone to error than would be someone with larger hands. Moreover, the keyboard is organized with the low notes to the left and the high notes to the right. This is an ideal arrangement for someone who is right handed because it enables most of the complicated passages to be played by the hand that is strongest, and most of the held chords to be played by the hand that is weakest. For someone who is left handed, however, it makes things more difficult. There were some new bits I wasn't sure about yet for which I would cross my hands and play the left part with the right hand and the right part with the left hand to more easily hear how it sounded. Only if I really liked what I heard did I go to the trouble of re-learning it with the parts switched so that my hands were not crossed.

A possible silver lining here is that because playing the piece provides constant reminders of the 'born into a universe where I don't belong' feeling that inspired it in the first place, I am less prone to disengaging from the fire that fuelled it, and therefore, I think, I play it with more feeling, and it stays fresh. Also, I suspect that lefthanders eventually develop the ability to create a very expressive sound simply because they are more forced to develop the fine control of their weaker hand to its full potential. Also, playing the bass notes with your stronger hand may have advantages. There are parts of the piece, and one in particular, with deep, booming chords played with the left hand and fast, intricate passages played with the right hand. Sure, it would be easier to play if I was wired the other way around. But I think there is a kind of power and confidence that comes through when someone is playing bass notes with their dominant hand, the hand that almost seems more connected to the core of who they are. It took some time to get this passage right, but the fact that I had to work at it increased the dramatic tension of witnessing its transition from something klunky and awkward sounding to something that, to me at least, was deeply moving. It is the closest thing I have experienced to 'giving birth'.

### Interpreting the Introspective Account in terms of Theories of Creativity

Since I am not a very accomplished pianist and have long forgotten the lessons I once took in theory and harmony, the process of composing this piece might have sounded to someone listening like a process of trial and error. Indeed, most standard theories of creativity would view these trials as independently generated entities competing with one another to be selected. It is widely assumed that the creative process involves *searching* through memory and/or *selecting* amongst a set of predefined candidate ideas. For example, computer scientists have modeled the creative process as heuristic search (e.g. Simon, 1973, 1986). In psychology, there is much evidence for, and discussion of, the role of *divergent thinking* in creativity (Guilford, 1968; for a review see Runco, 2010). Divergent thinking is said to involve the generation of multiple, often unconventional possibilities. Thus construed, is often thought of as going hand-in-hand with selection, since if you come up with multiple alternatives you eventually weed some of them out. Many well-known theories of creativity, such as the Geneplore model (Finke, Ward, & Smith, 1992), and the Darwinian theory of creativity (Simonton, 1999) involve two stages: generation of multiple possibilities, followed by exploration and ultimately selective retention of the most promising of them.

However, when I was composing 'Stream not gone dry', it did not feel like I was generating a collection of isolated possibilities and then selecting amongst them. It felt as if the piece had existed in my mind since the day it came to me as a sort of platonic essence, but it had not yet materialized in the external world. This is consistent with data on real-time studies of artists and designers indicate that creative ideation involves elaborating on a 'kernel idea', which goes from ill-defined to well-defined through an interaction between artist and artwork (Locher, 2010; Tovey & Porter, 2003; Weisberg, 2004).

The experience of composing this piece is in part what led me to propose that the generation stage of creative thinking is divergent not in the sense that it moves in multiple directions or generates multiple possibilities, but in the sense that it produces a raw idea that is vague or unfocused, that requires further processing to become viable (Gabora & Saab, 2010). Similarly, I have proposed that the exploration stage of creative thinking is *convergent* not in the sense that it entails selecting from amongst alternatives but in the sense that it entails considering a vague idea from different perspectives until it comes into focus. The idea, in other words, is that the terms divergent and convergent are applicable to creative thought not in the sense of going from one to many or from many to one, but in the sense of going from well-defined to ill-defined, and *vice versa*. Although a particular creative thinking process *may* involve search or selection amongst multiple possibilities, it *need* not, and selection need not figure prominently in a general theory of creativity (Gabora, 2005, 2010).

Ontologically, selection amongst multiple well-defined entities entails a different formal structure from actualizing the potential of a single, ill-defined entity (Gabora, 2005; Gabora & Aerts, 2005, 2007). Cognitively, thinking of a single vague idea seems relatively straightforward, whereas it is not obvious that one could simultaneously hold in one's mind multiple well-defined ideas. But perhaps the strongest reason to suppose that creativity involves, in the general case, not selection amongst multiple ideas but the honing of a half-baked idea, is that it is consistent with the structure of associative memory (Gabora, 2010; Gabora & Ranjan, 2013). Because memory is sparse, distributed, and content-addressable, knowledge and memories that are relevant to the situation or task at hand naturally come to mind (e.g. Hinton, McClelland, & Rumelhart, 1986; Kanerva, 1988). Neural cell assemblies that respond to the particular features of a situation are activated, and items previously encoded in these cell assemblies (that have similar constellations of features and activate similar distributed sets of neurons), are evoked.

Both the vagueness of a 'half-baked' idea and the sense that it holds potential, as well as its capacity to actualize in different ways depending on how one thinks it through, may be side effects of the phenomenon of interference. In interference, a recent memory interferes with the capacity to recall an older memory. A similar phenomenon occurs in neural networks, where it goes by different names: 'crosstalk', 'superposition catastrophe', 'false memories', 'spurious memories', and 'ghosts' (Feldman & Ballard, 1982; Hopfield, 1982; Hopfield, Feinstein, & Palmer, 1983). Interference is generally thought of as detrimental, but it may be of help with respect to creativity. A half-baked idea may be what results when two or more items encoded in overlapping distributions of neural cell assemblies interfere with each other and get evoked simultaneously.

The phenomenon of interference leading to creative ideation has been referred to as *creative interference* (Gabora & Saab, 2011; Ranjan & Gabora, 2012). When an idea emerges through creative interference it consists of multiple items merged into a single structure that is initially vague, 'half-baked', or *ill-defined*. The vagueness may reflect that it is uncertain how, in the context of each other, the elements that make up the idea come together as a realizable whole. This structure can be said to be in a *state of potentiality* because some features or elements could take on different values depending on how the idea unfolds. It is proposed that this unfolding involves disentangling the relevant features from the irrelevant features by observing how the idea looks from sequentially considered perspectives. In other words, one observes how it interacts with various contexts, either internally generated (think it through) or externally generated (try it out). Support for the hypothesis that midway through creative processing an idea is in a potentiality state comes from research on concept combination (Aerts & Gabora, 2005a,b) and analogy formation (Gabora & Saab, 2011).

If I allow myself to speak as I would to my artist friends, I would say that the state of 'Stream not gone dry', when it came to me, was complete and finished with respect to the reality it had come from, but in a state of potentiality with respect to the reality humans are most familiar with. When I read this over with my 'scientist' hat on, however, I have no clue what it means. I think that it is possible to describe a particular idea in a particular state as 'complete and whole' *and* to describe the same idea in the same state as 'vague and unfinished' and in both cases be convinced that you are telling the truth.

There is only one part of the piece that that did not emerge in a natural way from constraints already in place given what I had started with, and that required conscious work. I was in one key and I had to find a way of getting back to another key. Initially I had no clue how to proceed. I did try different possibilities in a way that could be described as 'generating variations until I found the solution to a problem'. Eventually I found a way of doing it that I was pleased with. It echoed back to the complex part that had originally been a separate piece of music. When this happened I experienced a feeling of recognition. At that point it no longer felt like I had been 'generating variations'; it felt as if I had lost the path and found it again.

## Conclusions

There are creative individuals who, after producing decent work, seem to descend into a state where everything they do comes across as a little narcissistic. I have held back from writing anything like this chapter before because I never wanted to be one of these people. But the longer you wait to tell a story, the blurrier the details become, until eventually they escape you. Part of why I wanted to write this out was just to achieve a sense of personal closure. 'Stream not gone dry' makes more sense in the context of not just the person I was when I conceived it, the circumstances I was in, and the times in which these circumstances occurred, as well as the specific thoughts and ideas I had while composing it. However, I sensed a widening gulf between who I am now, and the person who had particular experiences that seemed to be captured and resolved through the process of writing this music. Not only am I changing, but the world is changing, and I could feel the potential for the route back to become blurry. The other reason for writing this is scientific. I am convinced that introspective accounts do have a place in scholarly efforts to understand the creative process. They will never take the place of more controlled approaches, but I believe they have a vital role to play in the development of a comprehensive theoretical framework for creativity.

**Acknowledgments**

This work was funded by grants from the Natural Sciences and Engineering Research Council of Canada (NSERC), and the Concerted Research Program of the *Flemish Government of Belgium*.